\documentclass{article}

\usepackage{amssymb,amsfonts,amsmath}
\usepackage{cite,enumerate,float}
\usepackage{color}
\usepackage{pdflscape}

\hoffset= - 1.0in         

\begin{document}

\title{\vspace{0.1cm}{\Large {\bf  Racah matrices for the symmetric representation of the $SO(5)$ group.}\vspace{.2cm}}
\author{\bf Andrey Morozov$^{a,b,c}$\thanks{e-mail: morozov.andrey.a@iitp.ru}}
}
\date{ }

\maketitle

\vspace{-5.5cm}

\begin{center}
\hfill ITEP/TH-7/26\\
\hfill IITP/TH-7/26\\
\hfill MIPT/TH-7/26\\
\end{center}

\vspace{3.6cm}

\begin{center}
$^a$ {\small {\it IITP RAS, Moscow 127994, Russia}}\\
$^b$ {\small {\it MIPT, Dolgoprudny, 141701, Russia}}\\
$^c$ {\small {\it ITEP, Russia}}\\
\end{center}

\vspace{1cm}

\begin{abstract}
Approaches to calculate $SU(N)$ colored knot invariants (HOMFLY-PT polynomials) are well and widely developed. However, $SO(N)$ case is mostly forgotten. With this paper we want to start the discusion of how to generalize Reshetikhin-Turaev approach to the $SO(2n+1)$ case and which difficutlies arise in this discussion. We provide $\mathcal{R}$ and Racah matrices for the symmetric representation of the $SO(5)$ group and show how to find the corresponding Kauffmann polynomials.
\end{abstract}

\vspace{.5cm}

\section{Introduction}

Knot polynomials, their properties and applications are widely studied for the last thirty odd years. Especially widely studied was HOMFLY-PT polynomials, which are equal to the Wilson-loops in the $SU(N)$ Chern-Simons theory \cite{Witt}. Certain effective approaches for this case were developed \cite{RT,RT2,RT3,RTus,RTus2}, which use quantum $\mathcal{R}$-matrices and Racah matrices. These methods allow one to calculate colored HOMFLY-PT polynomials, i.e. polynomials corresponding to the larger highest-weight representations of the $SU(N)$ group. There are still certain unresolved problems with HOMFLY-PT polynomials, such as non-symmetric representations and Racah matrices with multiplicities \cite{Bish} or representations of quantum $SU(N)$ when $q$ is a root-of-unity \cite{ROU1,ROU2}.

However Wilson-loop averages and knot polynomials can be studied not only for the group $SU(N)$ but also for other groups. In the present paper we discuss the $SO(N)$ group. While the corresponding polynomials are known in literature and called Kauffmann polynomials, the same set of methods and approaches, as there is for the $SU(N)$, is virtually non-existent. Meanwhile these polynomials are different from the $SU(N)$ ones, representations are more complex. Also they are interesting in certain physical connections between physical models, such as topological strings, and knot theory \cite{Integr1,Integr}.

With this paper we want to start systematic description of modern Reshetikhin-Turaev approach and quantum Racah matrices for the $SO(N)$ group. There are some particular known cases, such as Rosso-Jones formula \cite{RJ1,RJ2,RJ3,Mkrt},  some answers for simple representations and knots and two-bridge knots in adjoint representation \cite{Mkrt,Eig6}. Polynomials for the $SO(3)$ group are equal to the Jones polynomials corresponding to the $SU(2)$ group. Therefore, we want to start with the next simplest case -- knot polynomials and corresponding Racah matrices and $\mathcal{R}$-matrices for the 3-strand knots in fundamental and symmetric representation of $SO(5)$ group.

The paper is structured as follows. In s.\ref{s:SuN} we remind the known approach to the calculus of the $SU(N)$, and in s.\ref{s:SoN} we discuss how to generalize it to the $SO(2n+1)$ case. In s.\ref{s:SoNfund} we show how it works for the fundamental representation of $SO(2n+1)$ and in s.\ref{s:so5} we discuss the symmetric representation of $SO(5)$ and what the $\mathcal{R}$ and Racah matrices look like. In s.\ref{s:Kpol} we calculate examples of Kauffmann polynomials for the symmetric representations of the $SO(5)$ group. In Appendix \ref{s:Mat} we provide the list of matrices for the symmetric representations of the $SO(5)$ group.

Throughout the paper we will use notations, common to the studies of the quantum group.
\begin{equation}
\{x\}\equiv x-x^{-1},\ \ \  [n]_q\equiv\cfrac{q^n-q^{-n}}{q-q^{-1}}, \ \ \  [\![n]\!]_q\equiv\cfrac{q^{n/2}-q^{-n/2}}{q^{1/2}-q^{-1/2}}.
\end{equation}
$[n]_q$ is called a quantum number, which is a common quantity in quantum groups. Also we will use the following notation for the Young diagrams -- non-increasing sequence of natural numbers:
\begin{equation}
Y=[\!\![\!\![y_1,y_2,\dots]\!\!]\!\!].
\end{equation}

\section{Basics of $SU(N)$ calculus \label{s:SuN}}

HOMFLY-PT polynomials and representations and Racah matrices for the $SU(N)$ case are well known. In this section we will describe the basics of these approaches so that then we can describe, how they will change for the $SO(2N+1)$ case. According to the modern Reshetikhin-Turaev approach, based on \cite{RT} and later developed in \cite{RT2,RT3,RTus,RTus2}, we discuss knot polynomial for a knot $\mathcal{K}$ in representation $T$ of the $SU(N)$ group as a character expansion\footnote{There is also a case of links which has several components which can carry different representations. This however is conceptually simple generalization, and we will omit this discussion in this paper.}
\begin{equation}
H^{\mathcal{K}}_{T}(A,q)=\sum\limits_{Q} S^*_Q(A,q) B^{\mathcal{K}}_Q(q).
\label{eq:Hcalc}
\end{equation}
For the $k$-strand braid and knot in representation $T$, $Q$ are irreducible representations appearing in the expansion
\begin{equation}
T^{\otimes k}=\sum Q,
\end{equation}
and $S^*_Q$ are Schur polynomials in special points, which are equal to the quantum dimensions of representation $Q$. Representations are usually labeled by the Young diagrams. It is important to note that\footnote{In fact it is also true if we take Schur polynomials themselves without specializing them to a special point. This is even more useful to find expansion into the irreducible representations.}
\begin{equation}
\left(S^*_T(A,q)\right)^k=\sum\limits_Q S^*_Q(A,q).
\end{equation}
This equation is useful to find expansion into irreducible representations.

$B_Q$ is a product of $\mathcal{R}$-matrices in the $k$-strand braid, corresponding to the particular knot we want to study. Quantum $\mathcal{R}$-matrix can be diagonalized, and its eigenvalues are quite simple. $\mathcal{R}$-matrix is an operator which acts on a pair of strands or, in other words, on a tensor product of a pair of representations. Due to the commutativity of coproduct and quantum $\mathcal{R}$-matrix, eigenvalues of the latter depend on irreducible representations in the product of these two representations:
\begin{equation}
T\otimes T=\sum Y,\ \ \ |\lambda_Y|\sim q^{\varkappa_Y+|Y|N/2-|Y|^2/2N}\equiv q^{a_Y},\ \ \ \varkappa_Y=\sum_{\{i,j\}\in Y} (i-j).
\end{equation}
Signs of the eigenvalues depend on whether representation $Y$ is in symmetric or antisymmetric square of representation \cite{sign,HW1,HW2}. The exact values of eigenvalues depend on the normalization of $\mathcal{R}$-matrix which should be chosen differently for different problems, these different normalization are called framings \cite{fr1,fr2,fr3,fr4}. The most important for the knot calculus are difference between eigenvalues for different representations which can be seen from this formula. Most used in the knot theory is the topological framing where different representations of the same knot has the same polynomial. In this case eigenvalues are equal to
\begin{equation}
|\lambda_Y|=q^{a_Y-4a_T}=q^{\varkappa_Y-4\varkappa_T-|T|N},
\end{equation}
since in the $SU(N)$ case $|Y|=2|T|$.

However, not all $\mathcal{R}$-matrices can be diagonalized at the same time if the braid has more than two strands. Transformation between bases where different $\mathcal{R}$-matrices are diagonal are described by the Racah matrices or 6-j symbols. These operators are defined for six representations. Namely, if we take three initial representations $T_1$, $T_2$ and $T_3$ then their product can be expanded in two different ways
\begin{equation}
\begin{array}{c}
(T_1\otimes T_2)\otimes T_3=\sum\limits_i Y_i\otimes T_3=\sum Q
\\
=
\\ \\
T_1\otimes (T_2\otimes T_3)=T_1\otimes \sum\limits_j Y_j=\sum Q.
\end{array}
\end{equation}
Transformation between these bases if we fix representation $Q$ at the end is given by the matrix
\begin{equation}
U_{Q}^{ij}=\left[
\begin{array}{lll}
T_1 & T_2 & Y_i
\\
T_3 & Q & Y_j
\end{array}\right].
\end{equation}
This matrix describes a rotation of the basis and, therefore, is an orthogonal matrix.

Since $\mathcal{R}$-matrices form a presentation of the braid group - they should satisfy Yang-Baxter equation:
\begin{equation}
\mathcal{R}_1\mathcal{R}_2\mathcal{R}_1=\mathcal{R}_2\mathcal{R}_1\mathcal{R}_2,
\end{equation}
or, using Racah matrices and diagonal $\mathcal{R}$-matrices,
\begin{equation}
\mathcal{R}U\mathcal{R}U^{\dagger}\mathcal{R}=U\mathcal{R}U^{\dagger}\mathcal{R}U\mathcal{R}U^{\dagger}.
\label{eq:Eig}
\end{equation}
This equation can be considered an equation on the diagonal $\mathcal{R}$-matrix and Racah matrix and it can be solved with respect to the Racah matrix at least for the matrices of small size \cite{EigIn}. This property is called Eigenvalue conjecture and allows us in many cases to get Racah matrices without using complex properties of quantum groups directly from the eigenvalues of $\mathcal{R}$-matrices. In the context of this paper we would need such equations for the matrices of the size 2-by-2:
\begin{equation}
U^{ii}=\frac{\sqrt{-\lambda_i\lambda_j}}{\lambda_i-\lambda_j},\ \
U^{ij}=\frac{\sqrt{\lambda_i^2-\lambda_i\lambda_j+\lambda_j^2}}{\lambda_i-\lambda_j},
\label{eq:2Rac}
\end{equation}
and 3-by-3
\begin{equation}
\begin{array}{l}
U^{ii}=\cfrac{-\lambda_i\sum_{k\neq i}\lambda_k}{\prod_{k\neq i}(\lambda_{i}-\lambda_{k})},
\\ \\
U^{ij}=\frac{\sqrt{-(\lambda_i^2+\lambda_j\lambda_k)(\lambda_j^2+\lambda_i\lambda_k)}}
{\sqrt{(\lambda_{i}-\lambda_{k})(\lambda_{j}-\lambda_{k})}(\lambda_i-\lambda_j)}.
\end{array}
\label{eq:3Rac}
\end{equation}

\section{$SO(2n+1)$ calculus \label{s:SoN}}

Now that we have described how the modern Reshetikhin-Turaev approach works for the $SU(N)$ case, we can discuss what is different for the $SO(2n+1)$ quantum groups. The main idea is still the same -- to use formula like (\ref{eq:Hcalc}):
\begin{equation}
K^{\mathcal{K}}_{T}(A,q)=\sum\limits_{Q} D_Q(A,q) B^{\mathcal{K}}_Q(q).
\label{KaufExp}
\end{equation}
Here $D_Q$ are now new quantum dimensions of the representations of quantum $SO(2n+1)$ group, which are different from $SU(N)$ \cite{Eig6,Mkrt,bfm}:
\begin{equation}
D_Q=\prod\limits_{1\leq i\leq j\leq l(Q)}\cfrac{\{q^{Q_i-Q_j+j-i}\}\{Aq^{Q_i+Q_j+1-i-j}\}}{\{q^{j-i}\}\{Aq^{1-i-j}\}}\times
\prod\limits_{k=1}^{l(Q)}\left(\cfrac{\{A^{1/2}q^{Q_k-k+1/2}\}}{\{A^{1/2}q^{-k+1/2}\}}
\prod_{s=1}^{Q_k}\cfrac{\{Aq^{Q_k+1-k-s-l(Q)}\}}{\{q^{s-k+l(Q)}\}}\right),
\end{equation}
where $l(Q)$ is a number of elements in the Young diagram $Q$, and $Q_i$ is the $i$-th element of the Young diagram $Q$. They can still be used to find expansion into the irreducible representations. The important difference is that for the $SU(N)$ case, when we looked at the expansion
\begin{equation}
T\otimes T=\sum Y,
\end{equation}
Young diagrams $Y$ encoding resulting representations has $|Y|$ boxes equal to twice the number of boxes $|T|$ (in general number of boxes is conserved when we sum representations and summed when we multiply them). However for the $SO(2n+1)$ case it is not true. For example for the simplest case of fundamental representation $[\!\![\!\![1]\!\!]\!\!]$:
\begin{equation}
[\!\![\!\![1]\!\!]\!\!]\otimes [\!\![\!\![1]\!\!]\!\!]=[\!\![\!\![2]\!\!]\!\!]+[\!\![\!\![1,1]\!\!]\!\!]+[\!\![\!\![0]\!\!]\!\!],
\end{equation}
where $[\!\![\!\![0]\!\!]\!\!]$ is trivial representation. This expansion is also known as
\begin{equation}
[\!\![\!\![1]\!\!]\!\!]\otimes [\!\![\!\![1]\!\!]\!\!]=\text{\textit{Symmetric}+\textit{Antisymmetric}+\textit{Trace}}.
\end{equation}
This difference actually leads to more significant changes further down the line.

The last point to mention here before moving to concrete examples is the change in $\mathcal{R}$-matrix eigenvalues:
\begin{equation}
T\otimes T=\sum Y,\ \ \ |\lambda_Y|\sim q^{\varkappa_Y+|Y|(N-1)/2}\equiv q^{b_Y},\ \ \ \varkappa_Y=\sum_{\{i,j\}\in Y} (i-j), \ \ N=2n+1.
\end{equation}
Also it is important to note that in this case $A=q^{N-1}=q^{2n}$.

\section{Fundamental representation of $SO(2n+1)$ \label{s:SoNfund}}

Let us start from the simple example of fundamental representation. As was already mentioned, an expansion of product of two fundamental representations is equal to
\begin{equation}
[\!\![\!\![1]\!\!]\!\!]\otimes[\!\![\!\![1]\!\!]\!\!]=[\!\![\!\![2]\!\!]\!\!]+[\!\![\!\![1,1]\!\!]\!\!]+[\!\![\!\![0]\!\!]\!\!].
\end{equation}
The corresponding eigenvalues in topological framing are equal to
\begin{equation}
\lambda_{[\!\![\!\![2]\!\!]\!\!]}=\cfrac{q}{A},\ \ \ \lambda_{[\!\![\!\![1,1]\!\!]\!\!]}=-\cfrac{1}{Aq},\ \ \ \lambda_{[\!\![\!\![0]\!\!]\!\!]}=\cfrac{1}{A^2}.
\label{eq:1eig}
\end{equation}
The first two eigenvalues are equal to the ones we have in the $SU(N)$ case, but the last one is different and, what is more important, we cannot renormalize the eigenvalues to get rid of $A$ which would have important consequence for the Racah matrices.

If we continue to multiply representations further to study $3$-strand braid, we will get
\begin{equation}
[\!\![\!\![1]\!\!]\!\!]\otimes [\!\![\!\![1]\!\!]\!\!]\otimes [\!\![\!\![1]\!\!]\!\!]= [\!\![\!\![3]\!\!]\!\!]+2[\!\![\!\![2,1]\!\!]\!\!]+[\!\![\!\![1,1,1]\!\!]\!\!]+3[\!\![\!\![1]\!\!]\!\!].
\end{equation}
Representations $[\!\![\!\![3]\!\!]\!\!]$ and $[\!\![\!\![1,1,1]\!\!]\!\!]$ appear only one time, therefore they will have Racah matrix of the size one-by-one and are equal to one. Representation $[\!\![\!\![2,1]\!\!]\!\!]$ appear two times. Therefore both $\mathcal{R}$-matrix and Racah matrix are of the size 2-by-2. Using (\ref{eq:1eig}) and (\ref{eq:2Rac}):
\begin{equation}
\mathcal{R}_{[\!\![\!\![2,1]\!\!]\!\!]}=\left(\begin{array}{cc} q/A \\ & -1/qA\end{array}\right),
\ \ \ \
U_{[\!\![\!\![2,1]\!\!]\!\!]}=\left(\begin{array}{cc} \cfrac{1}{[2]_q} & \cfrac{\sqrt{[3]_q}}{[2]_q} \\ \cfrac{\sqrt{[3]_q}}{[2]_q} & -\cfrac{1}{[2]_q}\end{array}\right).
\end{equation}
These matrices are in fact the same ones as appear in $SU(N)$ case. Representation $[\!\![\!\![1]\!\!]\!\!]$ appear three times and has the following matrices from  (\ref{eq:1eig}) and (\ref{eq:3Rac}):
\begin{footnotesize}
\begin{equation}
\begin{array}{c}
\mathcal{R}_{[\!\![\!\![1]\!\!]\!\!]}=\left(\begin{array}{ccc} q/A \\ & -1/qA \\ && 1\end{array}\right),
\\ \\
U_{[\!\![\!\![1]\!\!]\!\!]}=\left(\begin{array}{ccc} -\frac{[\![N-2]\!]_q[\![2]\!]_q}{[\![N]\!]_q[\![4]\!]_q} &
\frac{[\![2]\!]_q}{[\![4]\!]_q}\sqrt{\frac{[\![N-2]\!]_q[\![N+2]\!]_q[\![2N-8]\!]_q}{[\![N]\!]_q[\![N-4]\!]_q[\![2N-4]\!]_q}} &
-\frac{1}{[\![N]\!]_q}\sqrt{\frac{[\![2]\!]_q[\![N-2]\!]_q[\![N+2]\!]_q[\![2N-2]\!]_q}{[\![4]\!]_q[\![2N-4]\!]_q}}
\\ \frac{[\![2]\!]_q}{[\![4]\!]_q}\sqrt{\frac{[\![N-2]\!]_q[\![N+2]\!]_q[\![2N-8]\!]_q}{[\![N]\!]_q[\![N-4]\!]_q[\![2N-4]\!]_q}} & -\frac{[\![2N]\!]_q[\![N-2]\!]_q[\![2]\!]_q}{[\![N]\!]_q[\![2N-4]\!]_q[\![4]\!]_q} &
-\frac{[\![N-2]\!]_q}{[\![2N-4]\!]_q}\sqrt{\frac{[\![2]\!]_q[\![2N-2]\!]_q[\![2N-8]\!]_q}{[\![4]\!]_q[\![N]\!]_q[\![N-4]\!]_q}}
\\ -\frac{1}{[\![N]\!]_q}\sqrt{\frac{[\![2]\!]_q[\![N-2]\!]_q[\![N+2]\!]_q[\![2N-2]\!]_q}{[\![4]\!]_q[\![2N-4]\!]_q}} &
-\frac{[\![N-2]\!]_q}{[\![2N-4]\!]_q}\sqrt{\frac{[\![2]\!]_q[\![2N-2]\!]_q[\![2N-8]\!]_q}{[\![4]\!]_q[\![N]\!]_q[\![N-4]\!]_q}} &
-\frac{[\![N-2]\!]_q[\![2]\!]_q}{[\![N]\!]_q[\![2N-4]\!]_q}\end{array}\right).
\end{array}
\end{equation}
\end{footnotesize}

Due to the eigenvalues being of different degrees in $A$, Racah matrix depend on the $A$ and, therefore, on $n$, which did not happen in the $SU(N)$ case. It actually means that even if one manages to calculate Racah matrix directly for some value of $n$, it does not immediately give the answer for all the $SO(2n+1)$ groups. This makes approaches like the one described in \cite{HW1,HW2,HW3} for the $SU(N)$ case more difficult.

Dimensions of representations are equal to
\begin{equation}
\begin{array}{l}
D_{[\!\![\!\![1]\!\!]\!\!]}=\{A\}/\{q\}+1,\ \ \ D_{[\!\![\!\![3]\!\!]\!\!]}=\cfrac{\{A\}\{Aq\}\{Aq^2\}}{\{q\}\{q^2\}\{q^3\}}\left(\cfrac{\{q^3\}}{\{Aq^2\}}+1\right)
\\
\\
D_{[\!\![\!\![2,1]\!\!]\!\!]}=\cfrac{\{A\}\{Aq\}\{Aq^{-1}\}}{\{q\}^2\{q^3\}}\left(\cfrac{\{q^3\}}{\{A\}}+1\right),\ \ \
D_{[\!\![\!\![1,1,1]\!\!]\!\!]}=\cfrac{\{A\}\{Aq^{-1}\}\{Aq^{-2}\}}{\{q\}\{q^2\}\{q^3\}}\left(\cfrac{\{q^3\}}{\{Aq^{-2}\}}+1\right).
\end{array}
\end{equation}
Therefore we can find Kauffmann polynomials in fundamental representations for 3-strand knots. For example
\begin{equation}
K^{3_1}_{[\!\![\!\![1]\!\!]\!\!]}=\cfrac{q^2+q^{-2}}{A^2} + \cfrac{q-q^{-2}}{A^3}-\cfrac{q^2-1+q^{-2}}{A^4} - \cfrac{q-q^{-1}}{A^5},
\end{equation}
and
\begin{equation}
K^{4_1}_{[\!\![\!\![1]\!\!]\!\!]}=A^{2} (q^{2}-1+q^{-2})-A(q-q^{-1})(q^2-1+q^{-2})-(2q^2-3+2q^{-2})+\cfrac{(q-q^{-1})(q^2-1+q^{-2})}{A}+\cfrac{q^2-1+q^{-2}}{A^2}.
\end{equation}

\section{Symmetric representation of $SO(5)$\label{s:so5}}

Now let us move to the symmetric representation of $SO(2n+1)$ group. Again, expansion of product of two representations is equal to
\begin{equation}
  [\!\![\!\![2]\!\!]\!\!]\otimes [\!\![\!\![2]\!\!]\!\!]=[\!\![\!\![4]\!\!]\!\!]+[\!\![\!\![3,1]\!\!]\!\!]+[\!\![\!\![2,2]\!\!]\!\!]+[\!\![\!\![2]\!\!]\!\!]+[\!\![\!\![1,1]\!\!]\!\!]+[\!\![\!\![0]\!\!]\!\!]
\end{equation}
Therefore eigenvalues are equal to
\begin{equation}
\begin{array}{lllll}
\lambda_{[\!\![\!\![4]\!\!]\!\!]}=\cfrac{q^2}{A^2} & \ & \lambda_{[\!\![\!\![3,1]\!\!]\!\!]}=-\cfrac{1}{A^2q^2} & \ & \lambda_{[\!\![\!\![2,2]\!\!]\!\!]}=\cfrac{1}{A^2q^4}
\\
\ & \lambda_{[\!\![\!\![2]\!\!]\!\!]}=\cfrac{1}{A^3q^3} & \ & \lambda_{[\!\![\!\![1,1]\!\!]\!\!]}=-\cfrac{1}{A^3q^5}
\\
\ & \ & \lambda_{[\!\![\!\![0]\!\!]\!\!]}=\cfrac{1}{A^4q^4}
\end{array}
\label{2qubeig}
\end{equation}
Multiplying by the third symmetric representation we get
\begin{equation}
\begin{array}{l}
\ [\!\![\!\![4]\otimes [\!\![\!\![2]\!\!]\!\!]=[\!\![\!\![6]\!\!]\!\!]+[\!\![\!\![5,1]\!\!]\!\!]+[\!\![\!\![4,2]\!\!]\!\!]+[\!\![\!\![4]\!\!]\!\!]+[\!\![\!\![3,1]\!\!]\!\!]+[\!\![\!\![2]\!\!]\!\!],
\\
\ [\!\![\!\![3,1]\!\!]\!\!]\otimes [\!\![\!\![2]\!\!]\!\!]=[\!\![\!\![5,1]\!\!]\!\!]+[\!\![\!\![4,2]\!\!]\!\!]+[\!\![\!\![4,1,1]\!\!]\!\!]+[\!\![\!\![3,3]\!\!]\!\!]+[\!\![\!\![3,2,1]\!\!]\!\!]+[\!\![\!\![4]\!\!]\!\!]+2[\!\![\!\![3,1]\!\!]\!\!]+[\!\![\!\![2,2]\!\!]\!\!]+[\!\![\!\![2,1,1]\!\!]\!\!]+[\!\![\!\![2]\!\!]\!\!]+[\!\![\!\![1,1]\!\!]\!\!],
\\
\ [\!\![\!\![2,2]\!\!]\!\!]\otimes [\!\![\!\![2]\!\!]\!\!]=[\!\![\!\![4,2]\!\!]\!\!]+[\!\![\!\![3,2,1]\!\!]\!\!]+[\!\![\!\![2,2,2]\!\!]\!\!]+[\!\![\!\![3,1]\!\!]\!\!]+[\!\![\!\![2,2]\!\!]\!\!]+[\!\![\!\![2,1,1]\!\!]\!\!]+[\!\![\!\![2]\!\!]\!\!],
\\
\ [\!\![\!\![2]\!\!]\!\!]\otimes [\!\![\!\![2]\!\!]\!\!]=[\!\![\!\![4]\!\!]\!\!]+[\!\![\!\![3,1]\!\!]\!\!]+[\!\![\!\![2,2]\!\!]\!\!]+[\!\![\!\![2]\!\!]\!\!]+[\!\![\!\![1,1]\!\!]\!\!]+[\!\![\!\![0]\!\!]\!\!],
\\
\ [\!\![\!\![1,1]\!\!]\!\!]\otimes [\!\![\!\![2]\!\!]\!\!]=[\!\![\!\![3,1]\!\!]\!\!]+[\!\![\!\![2,1,1]\!\!]\!\!]+[\!\![\!\![2]\!\!]\!\!]+[\!\![\!\![1,1]\!\!]\!\!],
\\
\ [\!\![\!\![0]\!\!]\!\!]\otimes[\!\![\!\![2]\!\!]\!\!]=[\!\![\!\![2]\!\!]\!\!].
\end{array}
\label{2squarexp}
\end{equation}
And in total
\begin{equation}
\begin{array}{r}
[\!\![\!\![2]\!\!]\!\!]\otimes [\!\![\!\![2]\!\!]\!\!]\otimes [\!\![\!\![2]\!\!]\!\!]=[\!\![\!\![6]\!\!]\!\!]+2[\!\![\!\![5,1]\!\!]\!\!]+3[\!\![\!\![4,2]\!\!]\!\!]+[\!\![\!\![4,1,1]\!\!]\!\!]+[\!\![\!\![3,3]\!\!]\!\!]+2[\!\![\!\![3,2,1]\!\!]\!\!]+[\!\![\!\![2,2,2]\!\!]\!\!]+
\\ \\
+3[\!\![\!\![4]\!\!]\!\!]+6[\!\![\!\![3,1]\!\!]\!\!]+3[\!\![\!\![2,2]\!\!]\!\!]+3[\!\![\!\![2,1,1]\!\!]\!\!]+6[\!\![\!\![2]\!\!]\!\!]+3[\!\![\!\![1,1]\!\!]\!\!]+[\!\![\!\![0]\!\!]\!\!].
\end{array}
\label{2qubexp}
\end{equation}
\begin{landscape}
All the two-by-two and three-by-three matrices can be found using (\ref{eq:2Rac}) and (\ref{eq:3Rac}), we list them all in Appendix \ref{s:Mat}. The six-by-six matrix for representation $[\!\![\!\![2]\!\!]\!\!]$ can also be found using eigenvalue formula from \cite{Eig6}:


\setlength{\arraycolsep}{2pt}
\begin{equation}
\begin{array}{l}
U_{[\!\![\!\![2]\!\!]\!\!]}=
\\
\begin{footnotesize}
\left(
\begin{array}{cccccc}
\frac{[\![2]\!]_q[\![3]\!]_q[\![4]\!]_q[\![5]\!]_q}{[\![6]\!]_q[\![7]\!]_q[\![8]\!]_q[\![9]\!]_q} & \frac{[\![2]\!]_q[\![4]\!]_q[\![5]\!]_q}{[\![7]\!]_q[\![8]\!]_q}\sqrt{\frac{[\![3]\!]_q[\![11]\!]_q}{[\![6]\!]_q[\![9]\!]_q[\![10]\!]_q}} &
\frac{[\![2]\!]_q[\![4]\!]_q}{[\![6]\!]_q[\![8]\!]_q}\sqrt{\frac{[\![2]\!]_q[\![5]\!]_q[\![11]\!]_q}{[\![7]\!]_q[\![10]\!]_q}} & \frac{[\![4]\!]_q[\![5]\!]_q}{[\![9]\!]_q}\sqrt{\frac{[\![11]\!]_q[\![12]\!]_q}{[\![6]\!]_q[\![7]\!]_q[\![8]\!]_q[\![10]\!]_q}} & \frac{[\![4]\!]_q}{[\![6]\!]_q[\![7]\!]_q}\sqrt{\frac{[\![2]\!]_q[\![3]\!]_q[\![5]\!]_q[\![11]\!]_q[\![12]\!]_q}{[\![8]\!]_q[\![10]\!]_q}} & -\frac{\sqrt{[\![2]\!]_q[\![3]\!]_q[\![4]\!]_q[\![10]\!]_q[\![11]\!]_q[\![12]\!]_q}}{[\![6]\!]_q[\![7]\!]_q[\![8]\!]_q}
\\
\frac{[\![2]\!]_q[\![4]\!]_q[\![5]\!]_q}{[\![7]\!]_q[\![8]\!]_q}\sqrt{\frac{[\![3]\!]_q[\![11]\!]_q}{[\![6]\!]_q[\![9]\!]_q[\![10]\!]_q}} & \frac{[\![2]\!]^2_q[\![5]\!]_q[\![12]\!]_q}{[\![6]\!]_q[\![7]\!]_q[\![8]\!]_q[\![10]\!]_q}x_1 &
\frac{[\![4]\!]_q[\![5]\!]_q[\![14]\!]_q}{[\![7]\!]_q^2[\![8]\!]_q[\![10]\!]_q}\sqrt{\frac{[\![2]\!]_q[\![5]\!]_q[\![9]\!]_q}{[\![3]\!]_q[\![6]\!]_q[\![7]\!]_q}} & \frac{[\![2]\!]^2_q[\![5]\!]_q[\![14]\!]_q}{[\![7]\!]_q[\![10]\!]_q}\sqrt{\frac{[\![12]\!]_q}{[\![3]\!]_q[\![7]\!]_q[\![8]\!]_q[\![9]\!]_q}} & \frac{[\![2]\!]_q[\![3]\!]_q[\![5]\!]_q}{[\![7]\!]_q[\![10]\!]_q}\sqrt{\frac{[\![2]\!]_q[\![5]\!]_q[\![9]\!]_q[\![12]\!]_q}{[\![6]\!]_q[\![8]\!]_q}}x_5 & \frac{[\![3]\!]_q}{[\![7]\!]_q[\![8]\!]_q}\sqrt{\frac{[\![2]\!]_q[\![4]\!]_q[\![9]\!]_q}{[\![6]\!]_q}}
\\
\frac{[\![2]\!]_q[\![4]\!]_q}{[\![6]\!]_q[\![8]\!]_q}\sqrt{\frac{[\![2]\!]_q[\![5]\!]_q[\![11]\!]_q}{[\![7]\!]_q[\![10]\!]_q}} & \frac{[\![4]\!]_q[\![5]\!]_q[\![14]\!]_q}{[\![7]\!]_q^2[\![8]\!]_q[\![10]\!]_q}\sqrt{\frac{[\![2]\!]_q[\![5]\!]_q[\![9]\!]_q}{[\![3]\!]_q[\![6]\!]_q[\![7]\!]_q}} & \frac{[\![2]\!]^2_q[\![4]\!]_q[\![5]\!]_q}{[\![6]\!]_q[\![8]\!]_q[\![10]\!]_q}x_2 & -\frac{[\![2]\!]_q[\![5]\!]_q}{[\![3]\!]_q[\![10]\!]_q}\sqrt{\frac{[\![2]\!]_q[\![5]\!]_q[\![12]\!]_q}{[\![6]\!]_q[\![8]\!]_q}} & -\frac{[\![2]\!]_q\![5]\!]_q[\![14]\!]_q}{[\![6]\!]_q[\![7]\!]_q[\![10]\!]_q}\sqrt{\frac{[\![3]\!]_q[\![12]\!]_q}{[\![7]\!]_q[\![8]\!]_q}} & -\frac{[\![2]\!]_q}{[\![6]\!]_q[\![8]\!]_q}\sqrt{\frac{[\![3]\!]_q[\![4]\!]_q[\![5]\!]_q[\![12]\!]_q}{[\![7]\!]_q}}
\\
\frac{[\![4]\!]_q[\![5]\!]_q}{[\![9]\!]_q}\sqrt{\frac{[\![11]\!]_q[\![12]\!]_q}{[\![6]\!]_q[\![7]\!]_q[\![8]\!]_q[\![10]\!]_q}} & \frac{[\![2]\!]^2_q[\![5]\!]_q[\![14]\!]_q}{[\![7]\!]_q[\![10]\!]_q}\sqrt{\frac{[\![12]\!]_q}{[\![3]\!]_q[\![7]\!]_q[\![8]\!]_q[\![9]\!]_q}} & -\frac{[\![2]\!]_q[\![5]\!]_q}{[\![3]\!]_q[\![10]\!]_q}\sqrt{\frac{[\![2]\!]_q[\![5]\!]_q[\![12]\!]_q}{[\![6]\!]_q[\![8]\!]_q}} & \frac{[\![2]\!]^2_q[\![5]\!]_q}{[\![4]\!]_q[\![9]\!]_q[\![10]\!]_q}x_{3} & -\frac{[\![2]\!]_q[\![5]\!]_q[\![14]\!]_q}{[\![4]\!]_q[\![7]\!]_q[\![10]\!]_q}\sqrt{\frac{[\![2]\!]_q[\![3]\!]_q[\![5]\!]_q}{[\![6]\!]_q[\![7]\!]_q}} & -\sqrt{\frac{[\![2]\!]_q[\![3]\!]_q[\![4]\!]_q}{[\![6]\!]_q[\![7]\!]_q[\![8]\!]_q}}
\\
\frac{[\![4]\!]_q}{[\![6]\!]_q[\![7]\!]_q}\sqrt{\frac{[\![2]\!]_q[\![3]\!]_q[\![5]\!]_q[\![11]\!]_q[\![12]\!]_q}{[\![8]\!]_q[\![10]\!]_q}} & \frac{[\![2]\!]_q[\![3]\!]_q[\![5]\!]_q}{[\![7]\!]_q[\![10]\!]_q}\sqrt{\frac{[\![2]\!]_q[\![5]\!]_q[\![9]\!]_q[\![12]\!]_q}{[\![6]\!]_q[\![8]\!]_q}}x_5 & -\frac{[\![2]\!]_q\![5]\!]_q[\![14]\!]_q}{[\![6]\!]_q[\![7]\!]_q[\![10]\!]_q}\sqrt{\frac{[\![3]\!]_q[\![12]\!]_q}{[\![7]\!]_q[\![8]\!]_q}} &  -\frac{[\![2]\!]_q[\![5]\!]_q[\![14]\!]_q}{[\![4]\!]_q[\![7]\!]_q[\![10]\!]_q}\sqrt{\frac{[\![2]\!]_q[\![3]\!]_q[\![5]\!]_q}{[\![6]\!]_q[\![7]\!]_q}} & \frac{[\![2]\!]^3_q[3]\!]_q[\![5]\!]_q}{[\![4]\!]_q[\![6]\!]_q[\![7]\!]_q[\![10]\!]_q}x_{4} & \frac{[\![2]\!]_q[\![3]\!]_q}{[\![6]\!]_q[\![7]\!]_q}\sqrt{\frac{[\![4]\!]_q[\![5]\!]_q}{[\![8]\!]_q}}
\\
-\frac{\sqrt{[\![2]\!]_q[\![3]\!]_q[\![4]\!]_q[\![10]\!]_q[\![11]\!]_q[\![12]\!]_q}}{[\![6]\!]_q[\![7]\!]_q[\![8]\!]_q} & \frac{[\![3]\!]_q}{[\![7]\!]_q[\![8]\!]_q}\sqrt{\frac{[\![2]\!]_q[\![4]\!]_q[\![9]\!]_q}{[\![6]\!]_q}} & -\frac{[\![2]\!]_q}{[\![6]\!]_q[\![8]\!]_q}\sqrt{\frac{[\![3]\!]_q[\![4]\!]_q[\![5]\!]_q[\![12]\!]_q}{[\![7]\!]_q}} & -\sqrt{\frac{[\![2]\!]_q[\![3]\!]_q[\![4]\!]_q}{[\![6]\!]_q[\![7]\!]_q[\![8]\!]_q}} & \frac{[\![2]\!]_q[\![3]\!]_q}{[\![6]\!]_q[\![7]\!]_q}\sqrt{\frac{[\![4]\!]_q[\![5]\!]_q}{[\![8]\!]_q}} & \frac{[\![2]\!]_q[\![3]\!]_q[\![4]\!]_q}{[\![6]\!]_q[\![7]\!]_q[\![8]\!]_q}
\end{array}
\right)
\end{footnotesize}
\end{array}
\end{equation}
\setlength{\arraycolsep}{5pt}
where
\begin{equation}
\begin{array}{l}
x_{1}=q^3+2q+1+2q^{-1}+q^{-3}
\\
x_{2}=q^{6} - q^{5} + q^{4} - q^{3} + 2q^{2} - 2q + 3 - 2q^{-1} + 2q^{-2} - q^{-3} + q^{-4} - q^{-5} + q^{-6}
\\
x_{3}=q^{7}+q^{5}-q^{4}+q^{3}+3+q^{-3}-q^{-4}+q^{-5}+q^{-7}
\\
x_{4}=q^{5} + q^{3} + q^{2} + 1 + q^{-2} + q^{-3} + q^{-5}
\\
x_5=(\sqrt{q}-1/\sqrt{q})^4
\end{array}
\end{equation}

The remaining six-by-six matrix for representation $[\!\![\!\![3,1]\!\!]\!\!]$ however is more difficult to find. The equation (\ref{eq:Eig}) cannot be solved in a unique way if some eigenvalues coincide\footnote{In fact for every values of eigenvalues it has several solutions, however, if all eigenvalues are different, they differ only by signs of rows and columns.}. In this case in the sector corresponding to the coinciding eigenvalues Racah matrix can be rotated by any matrix, since they commute with $\mathcal{R}$-matrix. This is called a multiplicity case, since it happens when some representation appears two or more times in a product of two representations\footnote{In fact sometimes same eigenvalues can appear even for different representations, then Racah matrices have some interesting properties \cite{Bish}. However, it happens in the cases when there are also multiplicities.}.

Racah matrix for representation $[\!\![\!\![3,1]\!\!]\!\!]$ can be constructed by the modification of the highest-weight method for $SU(N)$, described in \cite{HW1,HW2,HW3}. However, exact details of modification of this method we will leave for \cite{SOwide}. The important difference from the $SU(N)$ case is that while for the $SU(N)$ we can calculate the answers directly for arbitrary value $N$, in this case we need to find separately for each group, and the higher the rank of the group is, the more difficult it is to build the highest weight vectors. However, we've managed to find this matrix for the $SO(5)$ group:
\setlength{\arraycolsep}{3pt}
\begin{equation}
\begin{array}{l}
U_{[\!\![\!\![3,1]\!\!]\!\!]}=
\\
\begin{footnotesize}
\left(
\begin{array}{cccccc}
\frac{[\![2]\!]_q[\![4]\!]_q[\![5]\!]_q}{[\![6]\!]_q[\![8]\!]_q[\![9]\!]_q} & -\frac{[\![4]\!]_q}{[\![8]\!]_q}\sqrt{\frac{[\![2]\!]_q[\![5]\!]_q[\![14]\!]_q}{[\![3]\!]_q[\![6]\!]_q[\![7]\!]_q[\![9]\!]_q[\![10]\!]_q}} &
\frac{[\![4]\!]_q^2}{[\![8]\!]_q}\sqrt{\frac{[\![5]\!]_q[\![11]\!]_q}{[\![3]\!]_q[\![6]\!]_q[\![9]\!]_q[\![10]\!]_q}} & \frac{[\![4]\!]_q}{[\![6]\!]_q}\sqrt{\frac{[\![2]\!]_q[\![5]\!]_q[\![11]\!]_q}{[\![3]\!]_q[\![9]\!]_q[\![10]\!]_q}} & \frac{[\![5]\!]_q}{[\![9]\!]_q}\sqrt{\frac{[\![2]\!]_q[\![4]\!]_q[\![11]\!]_q[\![14]\!]_q}{[\![3]\!]_q[\![6]\!]_q[\![8]\!]_q[\![10]\!]_q}} & -\sqrt{\frac{[\![2]\!]_q[\![5]\!]_q[\![11]\!]_q
[\![14]\!]_q}{[\![6]\!]_q[\![7]\!]_q[\![9]\!]_q[\![10]\!]_q}}
\\
-\frac{[\![4]\!]_q}{[\![8]\!]_q}\sqrt{\frac{[\![2]\!]_q[\![5]\!]_q[\![14]\!]_q}{[\![3]\!]_q[\![6]\!]_q[\![7]\!]_q[\![9]\!]_q[\![10]\!]_q}} & -\frac{[\![2]\!]_q[\![4]\!]_q}{[\![3]\!]_q[\![7]\!]_q[\![8]\!]_q[\![10]\!]_q}y_1 &
-\frac{[\![2]\!]_q[\![4]\!]_q[\![6]\!]_q}{[\![3]\!]_q^2[\![8]\!]_q[\![10]\!]_q}\sqrt{\frac{[\![2]\!]_q[\![11]\!]_q[\![14]\!]_q}{[\![7]\!]_q}} & -\frac{[\![2]\!]_q[\![5]\!]_q}{[\![3]\!]_q[\![10]\!]_q}\sqrt{\frac{[\![11]\!]_q[\![14]\!]_q}{[\![6]\!]_q[\![7]\!]_q}} & \frac{[\![2]\!]_q}{[\![3]\!]_q[\![10]\!]_q}\sqrt{\frac{[\![4]\!]_q[\![5]\!]_q[\![11]\!]_q}{[\![7]\!]_q[\![8]\!]_q[\![9]\!]_q}}y_2 & -\frac{[\![2]\!]_q[\![5]\!]_q}{[\![7]\!]_q[\![10]\!]_q}\sqrt{\frac{[\![11]\!]_q}{[\![3]\!]_q}}
\\
\frac{[\![4]\!]_q^2}{[\![8]\!]_q}\sqrt{\frac{[\![5]\!]_q[\![11]\!]_q}{[\![3]\!]_q[\![6]\!]_q[\![9]\!]_q[\![10]\!]_q}} & \frac{[\![2]\!]_q[\![4]\!]_q[\![6]\!]_q}{[\![3]\!]_q^2[\![8]\!]_q[\![10]\!]_q}\sqrt{\frac{[\![2]\!]_q[\![11]\!]_q[\![14]\!]_q}{[\![7]\!]_q}} & \frac{[\![2]\!]_q}{[\![3]\!]_q[\![7]\!]_q[\![8]\!]_q[\![10]\!]_q}y_3 & \frac{[\![2]\!]_q[\![5]\!]_q}{[\![3]\!]_q[\![4]\!]_q[\![10]\!]_q}\sqrt{\frac{[\![2]\!]_q}{[\![6]\!]_q}}y_4 & \frac{[\![2]\!]_q[\![14]\!]_q}{[\![3]\!]_q[\![7]\!]_q[\![10]\!]_q}\sqrt{\frac{[\![2]\!]_q[\![4]\!]_q[\![5]\!]_q[\![14]\!]_q}{[\![8]\!]_q[\![9]\!]_q}} & \frac{[\![5]\!]_q}{[\![10]\!]_q}\sqrt{\frac{[\![2]\!]_q[\![14]\!]_q}{[\![3]\!]_q[\![7]\!]_q}}
\\
-\frac{[\![4]\!]_q}{[\![6]\!]_q}\sqrt{\frac{[\![2]\!]_q[\![5]\!]_q[\![11]\!]_q}{[\![3]\!]_q[\![9]\!]_q[\![10]\!]_q}} & -\frac{[\![2]\!]_q[\![5]\!]_q}{[\![3]\!]_q[\![10]\!]_q}\sqrt{\frac{[\![11]\!]_q[\![14]\!]_q}{[\![6]\!]_q[\![7]\!]_q}} & -\frac{[\![2]\!]_q[\![5]\!]_q}{[\![3]\!]_q[\![4]\!]_q[\![10]\!]_q}\sqrt{\frac{[\![2]\!]_q}{[\![6]\!]_q}}y_4 & \frac{[\![2]\!]_q[\![2]\!]_q[\![5]\!]_q}{[\![3]\!]_q[\![4]\!]_q[\![6]\!]_q[\![10]\!]_q}y_{4} & \frac{[\![2]\!]_q[\![5]\!]_q}{[\![3]\!]_q[\![4]\!]_q[\![10]\!]_q}\sqrt{\frac{[\![4]\!]_q[\![5]\!]_q[\![8]\!]_q[\![14]\!]_q}{[\![6]\!]_q[\![9]\!]_q}} & \frac{[\![2]\!]_q[\![5]\!]_q}{[\![10]\!]_q}\sqrt{\frac{[\![14]\!]_q}{[\![3]\!]_q[\![6]\!]_q[\![7]\!]_q}}
\\
\frac{[\![5]\!]_q}{[\![9]\!]_q}\sqrt{\frac{[\![2]\!]_q[\![4]\!]_q[\![11]\!]_q[\![14]\!]_q}{[\![3]\!]_q[\![6]\!]_q[\![8]\!]_q[\![10]\!]_q}} & \frac{[\![2]\!]_q}{[\![3]\!]_q[\![10]\!]_q}\sqrt{\frac{[\![4]\!]_q[\![5]\!]_q[\![11]\!]_q}{[\![7]\!]_q[\![8]\!]_q[\![9]\!]_q}}y_2 & \frac{[\![2]\!]_q[\![14]\!]_q}{[\![3]\!]_q[\![7]\!]_q[\![10]\!]_q}\sqrt{\frac{[\![2]\!]_q[\![4]\!]_q[\![5]\!]_q[\![14]\!]_q}{[\![8]\!]_q[\![9]\!]_q}} &  -\frac{[\![2]\!]_q[\![5]\!]_q}{[\![3]\!]_q[\![10]\!]_q}\sqrt{\frac{[\![5]\!]_q[\![8]\!]_q[\![14]\!]_q}{[\![4]\!]_q[\![6]\!]_q[\![9]\!]_q}} & -\frac{[\![2]\!]_q[2]\!]_q[\![5]\!]_q[\![14]\!]_q}{[\![3]\!]_q[\![7]\!]_q[\![9]\!]_q[\![10]\!]_q} & -\frac{[\![2]\!]_q[\![5]\!]_q}{[\![10]\!]_q}\sqrt{\frac{[\![5]\!]_q[\![8]\!]_q}{[\![3]\!]_q[\![4]\!]_q[\![7]\!]_q[\![9]\!]_q}}
\\
-\sqrt{\frac{[\![2]\!]_q[\![5]\!]_q[\![11]\!]_q[\![14]\!]_q}{[\![6]\!]_q[\![7]\!]_q[\![9]\!]_q[\![10]\!]_q}} & -\frac{[\![2]\!]_q[\![5]\!]_q}{[\![7]\!]_q[\![10]\!]_q}\sqrt{\frac{[\![11]\!]_q}{[\![3]\!]_q}} & \frac{[\![5]\!]_q}{[\![10]\!]_q}\sqrt{\frac{[\![2]\!]_q[\![14]\!]_q}{[\![3]\!]_q[\![7]\!]_q}} & -\frac{[\![2]\!]_q[\![5]\!]_q}{[\![10]\!]_q}\sqrt{\frac{[\![14]\!]_q}{[\![3]\!]_q[\![6]\!]_q[\![7]\!]_q}} & \frac{[\![2]\!]_q[\![5]\!]_q}{[\![10]\!]_q}\sqrt{\frac{[\![5]\!]_q[\![8]\!]_q}{[\![3]\!]_q[\![4]\!]_q[\![7]\!]_q[\![9]\!]_q}} & -\frac{[\![2]\!]_q[\![5]\!]_q}{[\![7]\!]_q[\![10]\!]_q}
\end{array}
\right)
\end{footnotesize}
\end{array}
\end{equation}
where
\begin{equation}
\begin{array}{l}
y_{1}=q^{8}+q^{7}+2 q^{6}+3 q^{5}+4 q^{4}+4 q^{3}+6 q^{2}+6 q+7+6q^{-1}+6q^{-2}+4q^{-3}+4q^{-4}+3q^{-5}+2q^{-6}+q^{-7}+q^{-8}
\\
y_{2}=q^{5}+q^{3}+q^{2}+2 q+1+2q^{-1}+q^{-2}+q^{-3}+q^{-5}
\\
y_{3}=\sqrt(q)\left(2 q^{9}+5 q^{8}+9 q^{7}+14 q^{6}+19 q^{5}+23 q^{4}+26 q^{3}+27 q^{2}+28 q+29 +29q^{-1}+28q^{-2}+27q^{-3}+26q^{-4}+23q^{-5}+19q^{-6}+14q^{-7}+9q^{-8}+5q^{-9}+2q^{-10}\right)
\\
y_{4}=\frac{[\![18]\!]_q}{[\![9]\!]_q}-[\![4]\!]_q
\end{array}
\end{equation}
\setlength{\arraycolsep}{5pt}

Using the provided matrices one can find Kauffmann polynomials in symmetric representation using formula (\ref{KaufExp}).

\end{landscape}

\section{Kauffmann polynomials in symmetric representation \label{s:Kpol}}

To use matrices from the section \ref{s:so5} the first thing is to check some simple answers. According to (\ref{KaufExp}) polynomial of the unknot in any of its representations, see Fig.\ref{unknot}, should be equal to its quantum dimension
\begin{equation}
K_{[\!\![\!\![2]\!\!]\!\!]}^{o}=D_{[\!\![\!\![2]\!\!]\!\!]}=\frac{(A+q)(Aq^3-1)(A^2-1)q}{(q^2-1)(q^4-1)A^2}.
\end{equation}

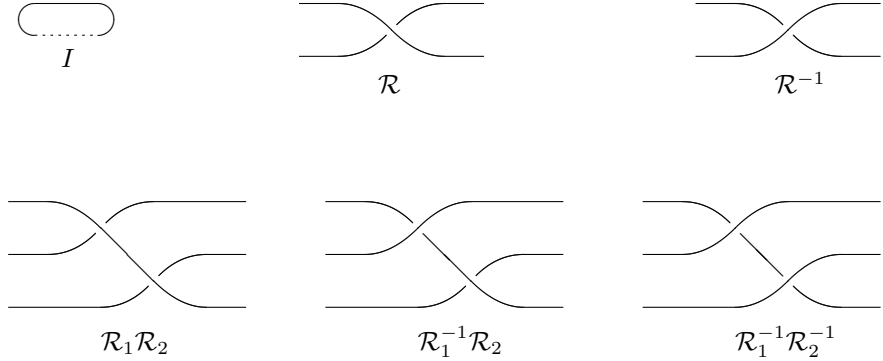
\begin{figure}[h!]
\begin{picture}(250,150)(-35,-125)
\put(50,20){
\put(0,0){\line(1,0){24}}
\put(0,-6){\oval(12,12)[l]}
\put(24,-6){\oval(12,12)[r]}
\multiput(0,-12)(3,0){8}{\line(1,0){1}}
\put(10,-24){\hbox{$I$}}
}
\put(150,0){
\put(0,20){\line(1,0){15}}
\put(0,0){\line(1,0){15}}
\qbezier(15,20)(25,20)(35,10)
\qbezier(15,0)(25,0)(33,8)
\qbezier(35,10)(45,0)(55,0)
\qbezier(37,12)(45,20)(55,20)
\put(55,20){\line(1,0){15}}
\put(55,0){\line(1,0){15}}
\put(30,-14){\hbox{$\mathcal{R}$}}
}
\put(300,0){
\put(0,20){\line(1,0){15}}
\put(0,0){\line(1,0){15}}
\qbezier(15,20)(25,20)(33,12)
\qbezier(15,0)(25,0)(35,10)
\qbezier(37,8)(45,0)(55,0)
\qbezier(35,10)(45,20)(55,20)
\put(55,20){\line(1,0){15}}
\put(55,0){\line(1,0){15}}
\put(30,-14){\hbox{$\mathcal{R}^{-1}$}}
}
\put(40,-75){
\put(0,20){\line(1,0){15}}
\put(0,0){\line(1,0){15}}
\put(0,-20){\line(1,0){35}}
\qbezier(15,20)(25,20)(35,10)
\qbezier(15,0)(25,0)(33,8)
\qbezier(37,12)(45,20)(55,20)
\put(35,10){\line(1,-1){20}}
\qbezier(35,-20)(45,-20)(53,-12)
\qbezier(55,-10)(65,-20)(75,-20)
\qbezier(57,-8)(65,0)(75,0)
\put(55,20){\line(1,0){35}}
\put(75,0){\line(1,0){15}}
\put(75,-20){\line(1,0){15}}
\put(35,-36){\hbox{$\mathcal{R}_1\mathcal{R}_2$}}
}
\put(160,-75){
\put(0,20){\line(1,0){15}}
\put(0,0){\line(1,0){15}}
\put(0,-20){\line(1,0){35}}
\qbezier(15,20)(25,20)(33,12)
\qbezier(15,0)(25,0)(35,10)
\qbezier(35,10)(45,20)(55,20)
\put(37,8){\line(1,-1){18}}
\qbezier(35,-20)(45,-20)(53,-12)
\qbezier(55,-10)(65,-20)(75,-20)
\qbezier(57,-8)(65,0)(75,0)
\put(55,20){\line(1,0){35}}
\put(75,0){\line(1,0){15}}
\put(75,-20){\line(1,0){15}}
\put(35,-36){\hbox{$\mathcal{R}^{-1}_1\mathcal{R}_2$}}
}
\put(280,-75){
\put(0,20){\line(1,0){15}}
\put(0,0){\line(1,0){15}}
\put(0,-20){\line(1,0){35}}
\qbezier(15,20)(25,20)(33,12)
\qbezier(15,0)(25,0)(35,10)
\qbezier(35,10)(45,20)(55,20)
\put(37,8){\line(1,-1){16}}
\qbezier(35,-20)(45,-20)(55,-10)
\qbezier(57,-12)(65,-20)(75,-20)
\qbezier(55,-10)(65,0)(75,0)
\put(55,20){\line(1,0){35}}
\put(75,0){\line(1,0){15}}
\put(75,-20){\line(1,0){15}}
\put(35,-36){\hbox{$\mathcal{R}^{-1}_1\mathcal{R}^{-1}_2$}}
}
\end{picture}
\caption{Some representations of unknot with one, two or three strands and corresponding $\mathcal{R}$-matrix expressions.}
\label{unknot}
\end{figure}
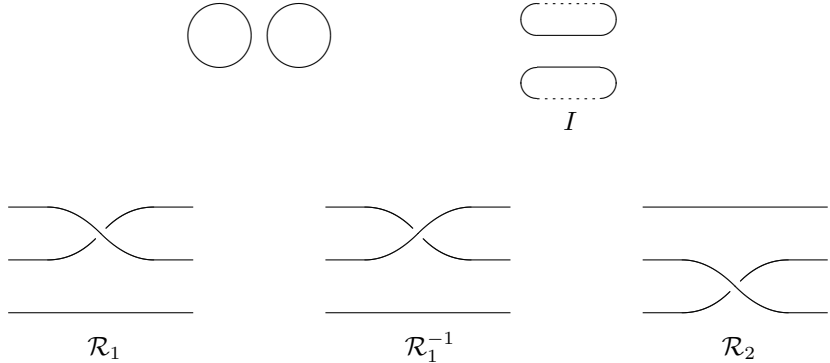
\begin{figure}[h!]
\begin{picture}(250,150)(-35,-125)
\put(120,10){
\put(0,0){\circle{24}}
\put(30,0){\circle{24}}
}
\put(240,10){
\put(0,0){\line(1,0){24}}
\put(0,6){\oval(12,12)[l]}
\put(24,6){\oval(12,12)[r]}
\multiput(0,12)(3,0){8}{\line(1,0){1}}
\put(0,-12){\line(1,0){24}}
\put(0,-18){\oval(12,12)[l]}
\put(24,-18){\oval(12,12)[r]}
\multiput(0,-24)(3,0){8}{\line(1,0){1}}
\put(10,-36){\hbox{$I$}}
}
\put(40,-75){
\put(0,20){\line(1,0){15}}
\put(0,0){\line(1,0){15}}
\put(0,-20){\line(1,0){70}}
\qbezier(15,20)(25,20)(35,10)
\qbezier(15,0)(25,0)(33,8)
\qbezier(37,12)(45,20)(55,20)
\qbezier(35,10)(45,0)(55,0)
\put(55,20){\line(1,0){15}}
\put(55,0){\line(1,0){15}}
\put(30,-36){\hbox{$\mathcal{R}_1$}}
}
\put(160,-75){
\put(0,20){\line(1,0){15}}
\put(0,0){\line(1,0){15}}
\put(0,-20){\line(1,0){70}}
\qbezier(15,20)(25,20)(33,12)
\qbezier(15,0)(25,0)(35,10)
\qbezier(35,10)(45,20)(55,20)
\qbezier(37,8)(45,0)(55,0)
\put(55,20){\line(1,0){15}}
\put(55,0){\line(1,0){15}}
\put(30,-36){\hbox{$\mathcal{R}^{-1}_1$}}
}
\put(280,-95){
\put(0,20){\line(1,0){15}}
\put(0,0){\line(1,0){15}}
\put(0,40){\line(1,0){70}}
\qbezier(15,20)(25,20)(35,10)
\qbezier(15,0)(25,0)(33,8)
\qbezier(37,12)(45,20)(55,20)
\qbezier(35,10)(45,0)(55,0)
\put(55,20){\line(1,0){15}}
\put(55,0){\line(1,0){15}}
\put(30,-16){\hbox{$\mathcal{R}_2$}}
}
\end{picture}
\caption{Some representations of two unentangled unknots with two or three strands and corresponding $\mathcal{R}$-matrix expressions.}
\label{2unknot}
\end{figure}

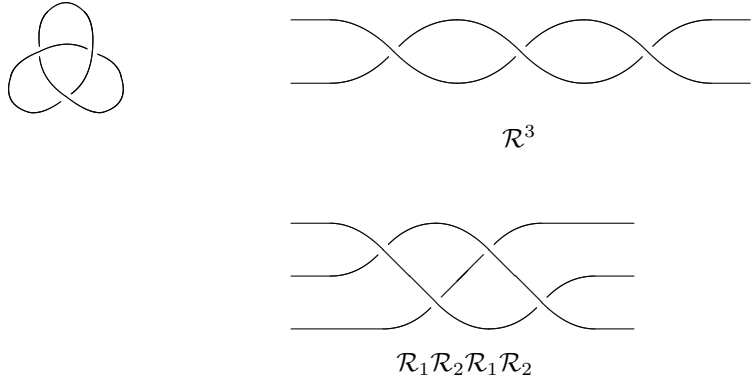
\begin{figure}[h!]
\begin{picture}(250,150)(-35,-125)
\put(80,10){
\qbezier(-10,0)(0,5)(8,1)
\qbezier(-10,-2)(-9,-11)(0,-17)
\qbezier(10,0)(9,-11)(2,-16)
\qbezier(10,0)(11,11)(7,15)
\qbezier(-10,2)(-11,11)(-7,15)
\qbezier(-7,15)(0,22)(7,15)
\qbezier(14,-23)(24,-20)(21,-10)
\qbezier(-14,-23)(-24,-20)(-21,-10)
\qbezier(12,-1)(20,-4)(21,-10)
\qbezier(-10,0)(-20,-4)(-21,-10)
\qbezier(0,-17)(9,-24)(14,-23)
\qbezier(-2,-19)(-9,-24)(-14,-23)
}
\put(165,10){
\put(0,12){\line(1,0){15}}
\put(0,-12){\line(1,0){15}}
\qbezier(15,12)(27,12)(39,0)
\qbezier(39,0)(51,-12)(63,-12)
\qbezier(15,-12)(27,-12)(37,-2)
\qbezier(41,2)(51,12)(63,12)
\qbezier(63,12)(75,12)(87,0)
\qbezier(87,0)(99,-12)(111,-12)
\qbezier(63,-12)(75,-12)(85,-2)
\qbezier(89,2)(99,12)(111,12)
\qbezier(111,12)(123,12)(135,0)
\qbezier(135,0)(147,-12)(159,-12)
\qbezier(111,-12)(123,-12)(133,-2)
\qbezier(137,2)(147,12)(159,12)
\put(159,12){\line(1,0){15}}
\put(159,-12){\line(1,0){15}}
\put(80,-36){\hbox{$\mathcal{R}^3$}}
}
\put(165,-75){
\put(0,20){\line(1,0){15}}
\put(0,0){\line(1,0){15}}
\put(0,-20){\line(1,0){35}}
\qbezier(15,20)(25,20)(35,10)
\qbezier(15,0)(25,0)(33,8)
\qbezier(37,12)(45,20)(55,20)
\put(35,10){\line(1,-1){20}}
\qbezier(35,-20)(45,-20)(53,-12)
\qbezier(55,-10)(65,-20)(75,-20)
\put(57,-8){\line(1,1){16}}
\qbezier(55,20)(65,20)(75,10)
\qbezier(77,12)(85,20)(95,20)
\put(75,10){\line(1,-1){20}}
\qbezier(75,-20)(85,-20)(93,-12)
\qbezier(95,-10)(105,-20)(115,-20)
\qbezier(97,-8)(105,0)(115,0)
\put(95,20){\line(1,0){35}}
\put(115,0){\line(1,0){15}}
\put(115,-20){\line(1,0){15}}
\put(40,-36){\hbox{$\mathcal{R}_1\mathcal{R}_2\mathcal{R}_1\mathcal{R}_2$}}
}
\end{picture}
\caption{Some representations of the trefoil knot with two or three strands and corresponding $\mathcal{R}$-matrix expressions.}
\label{trefoil}
\end{figure}

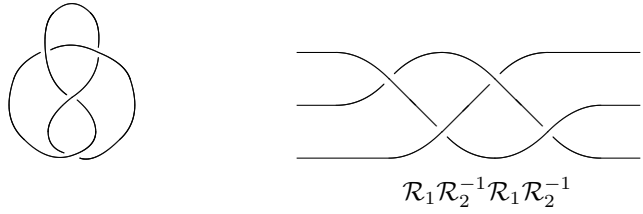
\begin{figure}[h!]
\begin{picture}(250,100)(-35,-50)
\put(80,10){
\qbezier(-8,1)(0,5)(10,0)
\qbezier(-10,0)(-9,-11)(-2,-16)
\qbezier(10,-2)(9,-11)(0,-17)
\qbezier(10,2)(11,11)(7,15)
\qbezier(-10,0)(-11,11)(-7,15)
\qbezier(-7,15)(0,22)(7,15)
\qbezier(10,0)(20,-4)(22,-10)
\qbezier(-12,0)(-20,-4)(-22,-10)
\qbezier(2,-18)(9,-24)(9,-30)
\qbezier(0,-17)(-9,-24)(-9,-30)
\qbezier(0,-39)(9,-36)(9,-30)
\qbezier(-3,-38)(-9,-36)(-9,-30)
\qbezier(3,-40)(9,-42)(18,-33)
\qbezier(22,-10)(27,-25)(18,-33)
\qbezier(-22,-10)(-27,-25)(-18,-33)
\qbezier(0,-39)(-9,-42)(-18,-33)
}
\put(165,-10){
\put(0,20){\line(1,0){15}}
\put(0,0){\line(1,0){15}}
\put(0,-20){\line(1,0){35}}
\qbezier(15,20)(25,20)(35,10)
\qbezier(15,0)(25,0)(33,8)
\qbezier(37,12)(45,20)(55,20)
\put(35,10){\line(1,-1){18}}
\qbezier(35,-20)(45,-20)(55,-10)
\qbezier(57,-12)(65,-20)(75,-20)
\put(55,-10){\line(1,1){18}}
\qbezier(55,20)(65,20)(75,10)
\qbezier(77,12)(85,20)(95,20)
\put(75,10){\line(1,-1){18}}
\qbezier(75,-20)(85,-20)(95,-10)
\qbezier(97,-12)(105,-20)(115,-20)
\qbezier(95,-10)(105,0)(115,0)
\put(95,20){\line(1,0){35}}
\put(115,0){\line(1,0){15}}
\put(115,-20){\line(1,0){15}}
\put(40,-36){\hbox{$\mathcal{R}_1\mathcal{R}^{-1}_2\mathcal{R}_1\mathcal{R}^{-1}_2$}}
}
\end{picture}
\caption{Figure-eight knot and its three-strand representation and corresponding $\mathcal{R}$-matrix expressions.}
\label{fig8}
\end{figure}

It can be checked that it is indeed the case for all the pictures. Another check can be done for a pair of unknots, see Fig. \ref{2unknot} where answer is equal to the square of quantum dimension
\begin{equation}
K_{[\!\![\!\![2]\!\!]\!\!]}^{o\times o}=D^2_{[\!\![\!\![2]\!\!]\!\!]}.
\end{equation}
Finally there is a trefoil knot, which is besides being a torus knot have both two-strand and three-strand representation, see Fig. \ref{trefoil}. In this case answer is equal to the
\begin{equation}
\begin{array}{r}
K^{3_1}_{[\!\![\!\![2]\!\!]\!\!]}=\frac{D_{[\!\![\!\![2]\!\!]\!\!]}}{q^{54}}(q^{42}+q^{36}+q^{35}+q^{34}-q^{32}-q^{31}+q^{29}+q^{28}-q^{26}-2 q^{25}-2 q^{24}-q^{23}+q^{22}+2 q^{21}+q^{20}-
\\ \\
-2 q^{18}-2 q^{17}-q^{16}+q^{15}+3 q^{14}+2 q^{13}-q^{11}-2 q^{10}-q^{9}+q^{8}+2 q^{7}+q^{6}-q^{4}-q^{3}-q^{2}+1)
\end{array}
\end{equation}
While using the approach, described in this paper, we can easily find polynomials for any 3-strand knot, let us provide one more example of a figure-eight knot, see Fig.\ref{fig8}. This is the simplest knot which is not 2-strand and not a torus knot. Its polynomial is equal to
\begin{equation}
\begin{array}{r}
K^{4_1}_{[\!\![\!\![2]\!\!]\!\!]}=\frac{D_{[\!\![\!\![2]\!\!]\!\!]}}{q^{28}}(q^{56}-q^{54}-q^{53}-q^{52}+2 q^{50}+3 q^{49}+q^{48}-2 q^{47}-4 q^{46}-3 q^{45}+5 q^{43}+7 q^{42}+2 q^{41}-4 q^{40}-
\\ \\
-7 q^{39}-7 q^{38}-q^{37}+7 q^{36}+10 q^{35}+5 q^{34}-4 q^{33}-10 q^{32}-9 q^{31}-2 q^{30}+7 q^{29}+13 q^{28}+
\\ \\
+7 q^{27}-2 q^{26}-9 q^{25}-10 q^{24}-4 q^{23}+5 q^{22}+10 q^{21}+7 q^{20}-q^{19}-7 q^{18}-7 q^{17}-4 q^{16}+
\\ \\
+2 q^{15}+7 q^{14}+5 q^{13}-3 q^{11}-4 q^{10}-2 q^{9}+q^{8}+3 q^{7}+2 q^{6}-q^{4}-q^{3}-q^{2}+1
\end{array}
\end{equation}

\section{Conclusion}

In this paper we discussed how th modern Reshsetikhin-Turaev approach used to calculate colored HOMFLY-PT polynomials should be modified for the $SO(2n+1)$ case and Kauufman polynomials. We discussed main differences, namely that Racah matrices now has direct dependence on the rank of the group and also multiplicities start to appear much earlier than in the $SU(N)$ case -- they appear already for the symmetric representation. As a consequence we were not able to find all the answers for the symmetric representation of all the $SO(2n+1)$ groups, and stopped only at the $SO(5)$. We checked the answer by calculating the corresponding Kauuffmann polynomials.

\section*{Acknowledgements}

This paper was supported by the state assignment of the Institute for Information Transmission
Problems of RAS.

We grateful for very useful discussions with A.Belov, L.Bishler, E.Lanina, A.Mironov, S.Mironov, N. Kolganov, A.Popolitov, M.Reva, A.Sleptsov, P.Suprun.

\appendix

\section{Racah and $\mathcal{R}$-matrices for the symmetric representation of $SO(5)$\label{s:Mat}}

Representations $[\!\![\!\![6]\!\!]\!\!]$, $[\!\![\!\![4,1,1]\!\!]\!\!]$, $[\!\![\!\![3,3]\!\!]\!\!]$, $[\!\![\!\![2,2,2]\!\!]\!\!]$ and $[\!\![\!\![0]\!\!]\!\!]$ appear only one time in (\ref{2qubexp}). Therefore, corresponding Racah matrices are equal to $1$:
\begin{equation}
U_{[\!\![\!\![6]\!\!]\!\!]}=U_{[\!\![\!\![4,1,1]\!\!]\!\!]}=U_{[\!\![\!\![3,3]\!\!]\!\!]}=U_{[\!\![\!\![2,2,2]\!\!]\!\!]}=U_{[\!\![\!\![0]\!\!]\!\!]}=\left(1\right).
\end{equation}
Corresponding $\mathcal{R}$-matrices can be written down from (\ref{2qubeig}):
\begin{equation}
\begin{array}{l}
\mathcal{R}_{[\!\![\!\![6]\!\!]\!\!]}=\left(\cfrac{q^2}{A^2}\right)=\left(\cfrac{1}{q^6}\right),\        \mathcal{R}_{[\!\![\!\![4,1,1]\!\!]\!\!]}=\mathcal{R}_{[\!\![\!\![3,3]\!\!]\!\!]}=\left(-\cfrac{1}{A^2q^2}\right)=\left(-\cfrac{1}{q^{10}}\right), \\ \\ \mathcal{R}_{[\!\![\!\![2,2,2]\!\!]\!\!]}=\left(\cfrac{1}{A^2q^4}\right)=\left(\cfrac{1}{q^{12}}\right),\ \mathcal{R}_{[\!\![\!\![0]\!\!]\!\!]}=\left(\cfrac{1}{A^3q^{3}}\right)=\left(\cfrac{1}{q^{15}}\right).
\end{array}
\end{equation}
Representations $[\!\![\!\![5,1]\!\!]\!\!]$ and $[\!\![\!\![3,2,1]\!\!]\!\!]$ appear two times in (\ref{2qubexp}). $\mathcal{R}$-matrices can be found from (\ref{2qubeig}) and (\ref{2squarexp}) and Racah matrices can be found using the Eigenvalue conjecture (\ref{eq:2Rac}):
\begin{equation}
\begin{array}{lclclclcl}
\mathcal{R}_{[\!\![\!\![5,1]\!\!]\!\!]}&=&\left(\begin{array}{cc}\frac{q^2}{A^2} \\ & -\frac{1}{A^2q^2}\end{array}\right)
&=&\left(\begin{array}{cc}\frac{1}{q^6} \\ & -\frac{1}{q^{10}}\end{array}\right),
&U_{[\!\![\!\![5,1]\!\!]\!\!]}&=&\left(\begin{array}{cc}\frac{[2]_q}{[4]_q} & \frac{\sqrt{[2]_q[6]_q}}{[4]_q}  \\ \frac{\sqrt{[2]_q[6]_q}}{[4]_q}  & -\frac{[2]_q}{[4]_q} \end{array}\right),
\\ \\
\mathcal{R}_{[\!\![\!\![3,2,1]\!\!]\!\!]}&=&\left(\begin{array}{cc}-\frac{1}{A^2q^2}\\ & \frac{1}{A^2q^4}\end{array}\right)
&=&\left(\begin{array}{cc}-\frac{1}{q^{10}}\\ & \frac{1}{q^{12}}\end{array}\right),
&
U_{[\!\![\!\![3,2,1]\!\!]\!\!]}&=&\left(\begin{array}{cc}\frac{1}{[2]_q} & \frac{\sqrt{[3]_q}}{[2]_q}  \\ \frac{\sqrt{[3]_q}}{[2]_q}  & -\frac{1}{[2]_q} \end{array}\right).
\end{array}
\end{equation}

\begin{landscape}

Representations $[\!\![\!\![4,2]\!\!]\!\!]$, $[\!\![\!\![4]\!\!]\!\!]$, $[\!\![\!\![2,2]\!\!]\!\!]$, $[\!\![\!\![2,1,1]\!\!]\!\!]$ and $[\!\![\!\![1,1]\!\!]\!\!]$ appear three times in (\ref{2qubexp}). $\mathcal{R}$-matrices can be found from (\ref{2qubeig}) and (\ref{2squarexp}) and Racah matrices can be found using the Eigenvalue conjecture (\ref{eq:3Rac}):
\begin{equation}
\begin{array}{l}
\mathcal{R}_{[\!\![\!\![4,2]\!\!]\!\!]}=\left(\begin{array}{ccc}\frac{q^2}{A^2} \\ & -\frac{1}{A^2q^2} \\ && \frac{1}{A^2q^4} \end{array}\right)
=\left(\begin{array}{ccc}\frac{1}{q^6} \\ & -\frac{1}{q^{10}} \\ && \frac{1}{q^{12}} \end{array}\right),
\ \ U_{[\!\![\!\![4,2]\!\!]\!\!]}=
\left(\begin{array}{ccc}-\frac{[2]_q}{[3]_q[4]_q} & \frac{[2]_q}{[4]_q}\sqrt{\frac{[5]_q}{[3]_q}} & \frac{\sqrt{[5]_q}}{[3]_q}
\\ \frac{[2]_q}{[4]_q}\sqrt{\frac{[5]_q}{[3]_q}} & -\frac{[6]_q}{[3]_q[4]_q} & \frac{1}{\sqrt{[3]_q}}
\\ \frac{\sqrt{[5]_q}}{[3]_q} &\frac{1}{\sqrt{[3]_q}}& -\frac{1}{[3]_q}
\end{array}\right),
\\ \\
\begin{array}{lcl}
\mathcal{R}_{[\!\![\!\![4]\!\!]\!\!]}&=&\left(\begin{array}{ccc}\frac{q^2}{A^2} \\ &-\frac{1}{A^2q^2}\\ & &\frac{1}{A^3q^3}\end{array}\right)
=\left(\begin{array}{ccc}\frac{1}{q^6}\\&-\frac{1}{q^{10}}\\ & & \frac{1}{q^{15}}\end{array}\right),
\\ \\
U_{[4]\!\!]\!\!]}&=&\left(\begin{array}{ccc}-\frac{[\![4]\!]_q[\![N]\!]_q}{[\![8]\!]_q[\![N+4]\!]_q} &
\frac{[\![4]\!]_q}{[\![8]\!]_q}\sqrt{\frac{[\![N]\!]_q[\![N+8]\!]_q[\![2N-8]\!]_q}{[\![2N]\!]_q[\![N+4]\!]_q[\![N-4]\!]_q}} &
\frac{1}{[\![N+4]\!]_q}\sqrt{\frac{[\![4]\!]_q[\![N]\!]_q[\![N+8]\!]_q[\![2N+4]\!]_q}{[\![8]\!]_q[\![2N]\!]_q}}
\\
\frac{[\![4]\!]_q}{[\![8]\!]_q}\sqrt{\frac{[\![N]\!]_q[\![N+8]\!]_q[\![2N-8]\!]_q}{[\![2N]\!]_q[\![N+4]\!]_q[\![N-4]\!]_q}} &
-\frac{[\![4]\!]_q[\![N]\!]_q[\![2N+8]\!]_q}{[\![8]\!]_q[\![2N]\!]_q[\![N+4]\!]_q} &
\frac{[\![N]\!]_q}{[\![2N]\!]_q}\sqrt{\frac{[\![4]\!]_q[\![2N+4]\!]_q[\![2N-8]\!]_q}{[\![8]\!]_q[\![N-4]\!]_q[\![N+4]\!]_q}}
\\
\frac{1}{[\![N+4]\!]_q}\sqrt{\frac{[\![4]\!]_q[\![N]\!]_q[\![N+8]\!]_q[\![2N+4]\!]_q}{[\![8]\!]_q[\![2N]\!]_q}}&
\frac{[\![N]\!]_q}{[\![2N]\!]_q}\sqrt{\frac{[\![4]\!]_q[\![2N+4]\!]_q[\![2N-8]\!]_q}{[\![8]\!]_q[\![N-4]\!]_q[\![N+4]\!]_q}}&
-\frac{[\![4]\!]_q[\![N]\!]_q}{[\![2N]\!]_q[\![N+4]\!]_q}
\end{array}\right)=
\\ \\
&=&\left(\begin{array}{ccc}-\frac{[\![4]\!]_q[\![5]\!]_q}{[\![8]\!]_q[\![9]\!]_q} &
\frac{[\![4]\!]_q}{[\![8]\!]_q}\sqrt{\frac{[\![2]\!]_q[\![5]\!]_q[\![13]\!]_q}{[\![9]\!]_q[\![10]\!]_q}} &
\frac{1}{[\![9]\!]_q}\sqrt{\frac{[\![4]\!]_q[\![5]\!]_q[\![13]\!]_q[\![14]\!]_q}{[\![8]\!]_q[\![10]\!]_q}}
\\
\frac{[\![4]\!]_q}{[\![8]\!]_q}\sqrt{\frac{[\![2]\!]_q[\![5]\!]_q[\![13]\!]_q}{[\![9]\!]_q[\![10]\!]_q}} &
-\frac{[\![4]\!]_q[\![5]\!]_q[\![18]\!]_q}{[\![8]\!]_q[\![9]\!]_q[\![10]\!]_q} &
\frac{[\![5]\!]_q}{[\![10]\!]_q}\sqrt{\frac{[\![2]\!]_q[\![4]\!]_q[\![14]\!]_q}{[\![8]\!]_q[\![9]\!]_q}}
\\
\frac{1}{[\![9]\!]_q}\sqrt{\frac{[\![4]\!]_q[\![5]\!]_q[\![13]\!]_q[\![14]\!]_q}{[\![8]\!]_q[\![10]\!]_q}}&
\frac{[\![5]\!]_q}{[\![10]\!]_q}\sqrt{\frac{[\![2]\!]_q[\![4]\!]_q[\![14]\!]_q}{[\![8]\!]_q[\![9]\!]_q}}&
-\frac{[\![4]\!]_q[\![5]\!]_q}{[\![9]\!]_q[\![10]\!]_q}
\end{array}\right),
\\ \\
\mathcal{R}_{[\!\![\!\![2,2]\!\!]\!\!]}&=&\left(\begin{array}{ccc}-\frac{1}{A^2q^2} \\ &\frac{1}{A^2q^4}\\ & &\frac{1}{A^3q^3}\end{array}\right)
=\left(\begin{array}{ccc}-\frac{1}{q^{10}}\\&\frac{1}{q^{12}}\\ & & \frac{1}{q^{15}}\end{array}\right),
\\ \\
U_{[\!\![\!\![2,2]\!\!]\!\!]}&=&\left(\begin{array}{ccc}-\frac{[\![2]\!]_q[\![N]\!]_q[\![2N-4]\!]_q}{[\![4]\!]_q[\![N-2]\!]_q[\![2N]\!]_q} &
\frac{[\![2]\!]_q}{[\![4]\!]_q}\sqrt{\frac{[\![N]\!]_q[\![N-4]\!]_q[\![2N+4]\!]_q}{[\![N-2]\!]_q[\![N+2]\!]_q[\![2N]\!]_q}} &
-\frac{[\![N]\!]_q}{[\![2N]\!]_q}\sqrt{\frac{[\![2]\!]_q[\![2N-2]\!]_q[\![2N+4]\!]_q}{[\![4]\!]_q[\![N-2]\!]_q[\![N+2]\!]_q}}
\\
\frac{[\![2]\!]_q}{[\![4]\!]_q}\sqrt{\frac{[\![N]\!]_q[\![N-4]\!]_q[\![2N+4]\!]_q}{[\![N-2]\!]_q[\![N+2]\!]_q[\![2N]\!]_q}} &
-\frac{[\![2]\!]_q[\![N]\!]_q}{[\![4]\!]_q[\![N-2]\!]_q} &
-\frac{1}{[\![N-2]\!]_q}\sqrt{\frac{[\![2]\!]_q[\![N]\!]_q[\![N-4]\!]_q[\![2N-2]\!]_q}{[\![4]\!]_q[\![2N]\!]_q}}
\\
-\frac{[\![N]\!]_q}{[\![2N]\!]_q}\sqrt{\frac{[\![2]\!]_q[\![2N-2]\!]_q[\![2N+4]\!]_q}{[\![4]\!]_q[\![N-2]\!]_q[\![N+2]\!]_q}}&
-\frac{1}{[\![N-2]\!]_q}\sqrt{\frac{[\![2]\!]_q[\![N]\!]_q[\![N-4]\!]_q[\![2N-2]\!]_q}{[\![4]\!]_q[\![2N]\!]_q}}&
\frac{[\![2]\!]_q[\![N]\!]_q}{[\![2N]\!]_q[\![N-2]\!]_q}
\end{array}\right)=
\\ \\
&=&\left(\begin{array}{ccc}-\frac{[\![2]\!]_q[\![5]\!]_q[\![6]\!]_q}{[\![3]\!]_q[\![4]\!]_q[\![10]\!]_q} &
\frac{[\![2]\!]_q}{[\![4]\!]_q}\sqrt{\frac{[\![5]\!]_q[\![14]\!]_q}{[\![3]\!]_q[\![7]\!]_q[\![10]\!]_q}} &
-\frac{[\![5]\!]_q}{[\![10]\!]_q}\sqrt{\frac{[\![2]\!]_q[\![8]\!]_q[\![14]\!]_q}{[\![3]\!]_q[\![4]\!]_q[\![7]\!]_q}}
\\
\frac{[\![2]\!]_q}{[\![4]\!]_q}\sqrt{\frac{[\![5]\!]_q[\![14]\!]_q}{[\![3]\!]_q[\![7]\!]_q[\![10]\!]_q}} &
-\frac{[\![2]\!]_q[\![5]\!]_q}{[\![3]\!]_q[\![4]\!]_q} &
-\frac{1}{[\![3]\!]_q}\sqrt{\frac{[\![2]\!]_q[\![5]\!]_q[\![8]\!]_q}{[\![4]\!]_q[\![10]\!]_q}}
\\
-\frac{[\![5]\!]_q}{[\![10]\!]_q}\sqrt{\frac{[\![2]\!]_q[\![8]\!]_q[\![14]\!]_q}{[\![3]\!]_q[\![4]\!]_q[\![7]\!]_q}}&
-\frac{1}{[\![3]\!]_q}\sqrt{\frac{[\![2]\!]_q[\![5]\!]_q[\![8]\!]_q}{[\![4]\!]_q[\![10]\!]_q}}&
\frac{[\![2]\!]_q[\![5]\!]_q}{[\![3]\!]_q[\![10]\!]_q}
\end{array}\right),
\end{array}
\end{array}
\end{equation}
\end{landscape}
\begin{equation*}
\begin{array}{lcl}
\mathcal{R}_{[\!\![\!\![2,1,1]\!\!]\!\!]}&=&\left(\begin{array}{ccc}-\frac{1}{A^2q^2} \\ &\frac{1}{A^2q^4}\\ & &-\frac{1}{A^3q^5}\end{array}\right)
=\left(\begin{array}{ccc}-\frac{1}{q^{10}}\\&\frac{1}{q^{12}}\\ & & -\frac{1}{q^{17}}\end{array}\right),
\\ \\
U_{[\!\![\!\![2,1,1]\!\!]\!\!]}&=&\left(\begin{array}{ccc}-\frac{[\![2]\!]_q[\![N]\!]_q}{[\![4]\!]_q[\![N+2]\!]_q} &
\frac{[\![2]\!]_q}{[\![4]\!]_q}\sqrt{\frac{[\![N]\!]_q[\![N+4]\!]_q[\![2N-4]\!]_q}{[\![N-2]\!]_q[\![N+2]\!]_q[\![2N]\!]_q}} &
-\frac{1}{[\![N+2]\!]_q}\sqrt{\frac{[\![2]\!]_q[\![N]\!]_q[\![N+4]\!]_q[\![2N+2]\!]_q}{[\![4]\!]_q[\![2N]\!]_q}}
\\
\frac{[\![2]\!]_q}{[\![4]\!]_q}\sqrt{\frac{[\![N]\!]_q[\![N+4]\!]_q[\![2N-4]\!]_q}{[\![N-2]\!]_q[\![N+2]\!]_q[\![2N]\!]_q}} &
-\frac{[\![2]\!]_q[\![N]\!]_q[\![2N+4]\!]_q}{[\![4]\!]_q[\![N+2]\!]_q[\![2N]\!]_q} &
-\frac{[\![N]\!]_q}{[\![2N]\!]_q}\sqrt{\frac{[\![2]\!]_q[\![2N+2]\!]_q[\![2N-4]\!]_q}{[\![4]\!]_q[\![N+2]\!]_q[\![N-2]\!]_q}}
\\
-\frac{1}{[\![N+2]\!]_q}\sqrt{\frac{[\![2]\!]_q[\![N]\!]_q[\![N+4]\!]_q[\![2N+2]\!]_q}{[\![4]\!]_q[\![2N]\!]_q}}&
-\frac{[\![N]\!]_q}{[\![2N]\!]_q}\sqrt{\frac{[\![2]\!]_q[\![2N+2]\!]_q[\![2N-4]\!]_q}{[\![4]\!]_q[\![N+2]\!]_q[\![N-2]\!]_q}}&
-\frac{[\![2]\!]_q[\![N]\!]_q}{[\![2N]\!]_q[\![N+2]\!]_q},
\end{array}\right)=
\\ \\
&=&\left(\begin{array}{ccc}-\frac{[\![2]\!]_q[\![5]\!]_q}{[\![4]\!]_q[\![7]\!]_q} &
\frac{[\![2]\!]_q}{[\![4]\!]_q}\sqrt{\frac{[\![5]\!]_q[\![6]\!]_q[\![9]\!]_q}{[\![3]\!]_q[\![7]\!]_q[\![10]\!]_q}} &
-\frac{1}{[\![7]\!]_q}\sqrt{\frac{[\![2]\!]_q[\![5]\!]_q[\![9]\!]_q[\![12]\!]_q}{[\![4]\!]_q[\![10]\!]_q}}
\\
\frac{[\![2]\!]_q}{[\![4]\!]_q}\sqrt{\frac{[\![5]\!]_q[\![6]\!]_q[\![9]\!]_q}{[\![3]\!]_q[\![7]\!]_q[\![10]\!]_q}} &
-\frac{[\![2]\!]_q[\![5]\!]_q[\![14]\!]_q}{[\![4]\!]_q[\![7]\!]_q[\![10]\!]_q} &
-\frac{[\![5]\!]_q}{[\![10]\!]_q}\sqrt{\frac{[\![2]\!]_q[\![6]\!]_q[\![12]\!]_q}{[\![3]\!]_q[\![4]\!]_q[\![7]\!]_q}}
\\
-\frac{1}{[\![7]\!]_q}\sqrt{\frac{[\![2]\!]_q[\![5]\!]_q[\![9]\!]_q[\![12]\!]_q}{[\![4]\!]_q[\![10]\!]_q}}&
-\frac{[\![5]\!]_q}{[\![10]\!]_q}\sqrt{\frac{[\![2]\!]_q[\![6]\!]_q[\![12]\!]_q}{[\![3]\!]_q[\![4]\!]_q[\![7]\!]_q}}&
-\frac{[\![2]\!]_q[\![5]\!]_q}{[\![7]\!]_q[\![10]\!]_q},
\end{array}\right),
\\ \\
\mathcal{R}_{[\!\![\!\![1,1]\!\!]\!\!]}&=&\left(\begin{array}{ccc}-\frac{1}{A^2q^2} \\ &\frac{1}{A^3q^3}\\ & &-\frac{1}{A^3q^5}\end{array}\right)
=\left(\begin{array}{ccc}-\frac{1}{q^{10}}\\&\frac{1}{q^{15}}\\ & & -\frac{1}{q^{17}}\end{array}\right),
\\ \\
U_{[\!\![\!\![1,1]\!\!]\!\!]}&=&\left(\begin{array}{ccc}-\frac{[\![2]\!]_q[\![N]\!]_q}{[\![2N]\!]_q[\![N+2]\!]_q} &
\frac{[\![N]\!]_q}{[\![2N]\!]_q}\sqrt{\frac{[\![2]\!]_q[\![2N+2]\!]_q[\![2N-4]\!]_q}{[\![4]\!]_q[\![N+2]\!]_q[\![N-2]\!]_q}} &
-\frac{1}{[\![N+2]\!]_q}\sqrt{\frac{[\![2]\!]_q[\![N]\!]_q[\![N+4]\!]_q[\![2N+2]\!]_q}{[\![4]\!]_q[\![2N]\!]_q}}
\\
\frac{[\![N]\!]_q}{[\![2N]\!]_q}\sqrt{\frac{[\![2]\!]_q[\![2N+2]\!]_q[\![2N-4]\!]_q}{[\![4]\!]_q[\![N+2]\!]_q[\![N-2]\!]_q}} &
-\frac{[\![2]\!]_q[\![N]\!]_q[\![2N+4]\!]_q}{[\![4]\!]_q[\![N+2]\!]_q[\![2N]\!]_q} &
-\frac{[\![2]\!]_q}{[\![4]\!]_q}\sqrt{\frac{[\![N]\!]_q[\![N+4]\!]_q[\![2N-4]\!]_q}{[\![N+2]\!]_q[\![N-2]\!]_q[\![2N]\!]_q}}
\\
-\frac{1}{[\![N+2]\!]_q}\sqrt{\frac{[\![2]\!]_q[\![N]\!]_q[\![N+4]\!]_q[\![2N+2]\!]_q}{[\![4]\!]_q[\![2N]\!]_q}}&
-\frac{[\![2]\!]_q}{[\![4]\!]_q}\sqrt{\frac{[\![N]\!]_q[\![N+4]\!]_q[\![2N-4]\!]_q}{[\![N+2]\!]_q[\![N-2]\!]_q[\![2N]\!]_q}}&
-\frac{[\![2]\!]_q[\![N]\!]_q}{[\![4]\!]_q[\![N+2]\!]_q}
\end{array}\right)=
\\ \\
&=&\left(\begin{array}{ccc}-\frac{[\![2]\!]_q[\![5]\!]_q}{[\![7]\!]_q[\![10]\!]_q} &
\frac{[\![5]\!]_q}{[\![10]\!]_q}\sqrt{\frac{[\![2]\!]_q[\![6]\!]_q[\![12]\!]_q}{[\![3]\!]_q[\![4]\!]_q[\![7]\!]_q}} &
-\frac{1}{[\![7]\!]_q}\sqrt{\frac{[\![2]\!]_q[\![5]\!]_q[\![9]\!]_q[\![12]\!]_q}{[\![4]\!]_q[\![10]\!]_q}}
\\
\frac{[\![5]\!]_q}{[\![10]\!]_q}\sqrt{\frac{[\![2]\!]_q[\![6]\!]_q[\![12]\!]_q}{[\![3]\!]_q[\![4]\!]_q[\![7]\!]_q}} &
-\frac{[\![2]\!]_q[\![5]\!]_q[\![14]\!]_q}{[\![4]\!]_q[\![7]\!]_q[\![10]\!]_q} &
-\frac{[\![2]\!]_q}{[\![4]\!]_q}\sqrt{\frac{[\![5]\!]_q[\![6]\!]_q[\![9]\!]_q}{[\![3]\!]_q[\![7]\!]_q[\![10]\!]_q}}
\\
-\frac{1}{[\![7]\!]_q}\sqrt{\frac{[\![2]\!]_q[\![5]\!]_q[\![9]\!]_q[\![12]\!]_q}{[\![4]\!]_q[\![10]\!]_q}}&
-\frac{[\![2]\!]_q}{[\![4]\!]_q}\sqrt{\frac{[\![5]\!]_q[\![6]\!]_q[\![9]\!]_q}{[\![3]\!]_q[\![7]\!]_q[\![10]\!]_q}}&
-\frac{[\![2]\!]_q[\![5]\!]_q}{[\![4]\!]_q[\![7]\!]_q}
\end{array}\right),
\end{array}
\end{equation*}

\end{document}